\documentclass[%
 reprint,
 amsmath,amssymb,
 aps,
]{revtex4-2}

\usepackage{graphicx}
\usepackage{dcolumn}
\usepackage{bm}
\usepackage[dvipsnames]{xcolor}
\usepackage{url}
\usepackage{soul}

\usepackage{todonotes}

\begin{document}


\title{{Statistical Dependencies Beyond Linear Correlations in Light Scattered by} Disordered Media}

\author{Ilya Starshynov}
\email{ilya.starshynov@glasgow.ac.uk}
 
\author{Alex Turpin}%
\author{Philip Binner}%
\author{Daniele Faccio}%
\email{daniele.faccio@glasgow.ac.uk}
\affiliation{School of Physics \& Astronomy, University of Glasgow, Glasgow G12 8QQ, United Kingdom }

\date{\today}

\begin{abstract}
Imaging through scattering and random media is an outstanding problem that to date has been tackled by either measuring the medium transmission matrix or exploiting linear correlations in the transmitted speckle patterns. However, {transmission matrix} techniques require interferometric stability  and linear correlations, such as the memory effect, can be exploited only in thin scattering media. Here we show the existence of a statistical dependency in strongly scattered optical fields in a case where first order correlations are not expected. We also show that this statistical dependence and the related  information transport is directly linked to  artificial neural network imaging in strongly scattering, dynamic media. These non-trivial dependencies provide a key to imaging through dynamic and thick scattering media with applications for deep-tissue imaging or imaging through  smoke or fog.

\end{abstract}

\maketitle
{\bf{Introduction.}} 
The statistics of individual speckle patterns created by coherent illumination of an optically random medium is well understood. It is widely accepted that as the real and imaginary parts of the scattered field are uncorrelated Gaussian random variables, the field amplitude follows a Rayleigh distribution and the light intensity follows a negative exponential distribution \cite{goodman2007speckle}. Identifying the correlation statistics of scattered light is a more challenging problem. Various types of correlations in space, time or frequency may exist in the scattered fields, depending on the problem geometry and disorder strength~\cite{Berkovitsa,Feng1988}. A particularly relevant configuration involves imaging through a scattering slab, where a common goal is to reconstruct the image of an object from the transmitted scattered light. The strongest and most evident of the field correlations, the memory effect (ME)~\cite{freund_memory_1988,Feng1988}, has proven to be useful in such a situation, for example allowing to reconstruct the image from a simple autocorrelation calculation of the speckle pattern~\cite{freund_looking_1990,bertolotti_non-invasive_2012,katz_non-invasive_2014}. Recently more subtle, long-range mesoscopic correlations emerging in strongly scattering media~\cite{starshynov2018non} have been used to retrieve the image of a hidden object~\cite{paniagua2019blind}. However, first order correlations capture only a fraction of the total statistical dependence between random variables~\cite{anderson2003an}.

\begin{figure}[t!]
    \includegraphics[width=8cm]{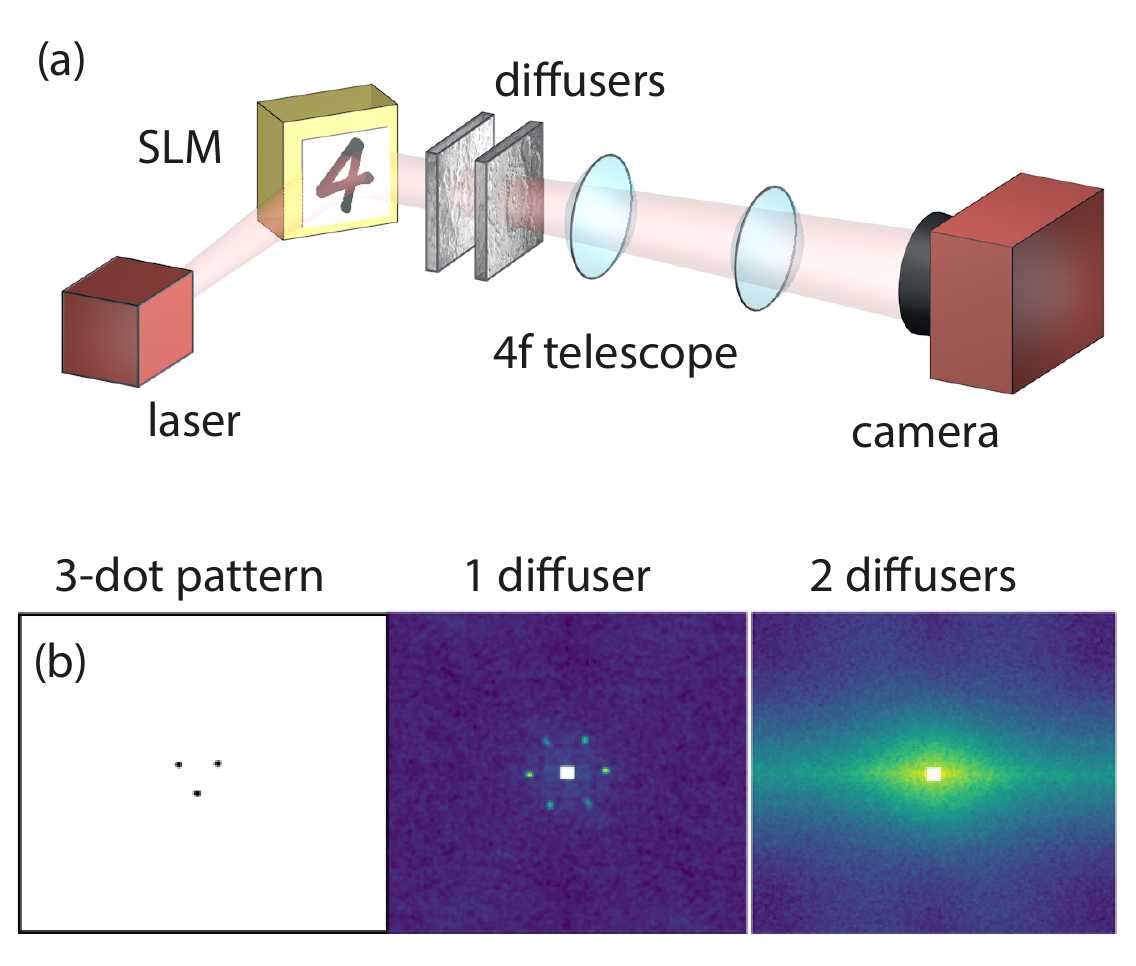}
    \caption{{\bf{Experimental and neural network layout}} (a) Experimental setup: a laser illuminates a {SLM  (DMD or liquid crystal)} which is imaged onto the camera (CCD). The scattering medium is made from one or two glass diffusers {separated by 5mm}. (b) A 3-dot pattern projected by the DMD was used to test for the presence of the memory effect (the separation between the dots was 0.8 mm). With 1 diffuser, the averaged output speckle autocorrelation clearly shows correlation features, indicating the presence of the ME. With 2 diffusers, the correlation pattern completely disappears, indicating the absence of any ME. Central peak with $C$ = 1 is removed for better visibility. }
    \label{fig:one}
\end{figure}

\indent The dependence between random variables can be embedded into higher order correlations even when the first order correlation is zero. As quantifying all the higher order correlations can be challenging especially for the case of multiple variables, usually information-theoretical criteria are applied to analyze a statistical dependence. In particular non-zero mutual information (MI) implies a dependence between the random variables. Information theory approach  has been applied to wave scattering in the context of radio-wave communication~\cite{foschini1998limits}, albeit focusing more on temporal or frequency modulations rather than spatial information~\cite{moustakas2000communication,simon2001communication,staring2004fluctuations}. In recent work \cite{byrnes_universal_2020}, the universal bounds of spatial information preserved in multiple scattering have been estimated within the context of random matrix theory~\cite{beenakker1997random}, thus neglecting the details of a realistic optical random scattering potential. \\
\indent {Alternatively to a stochastic approach to the characterisation of multiple scattering media, it is possible to formulate a deterministic description based on a transmission matrix (TM) measurement~\cite{Popoff2010,van2010information}.} This complex-valued matrix completely characterises the mapping between the input and output fields and once known, can be used to {calculate the input field distribution} from the output speckle pattern. The TM approach has been successfully {applied} to both diffuse imaging and imaging through multimode fibres~\cite{popoff_image_2010,ploschner2015seeing}. However, it suffers from the main drawback that it is sensitive to optical wavelength scale changes in the scattering medium. Measuring the transmission matrix typically requires some form of holography that needs to be repeated for a number of input modes and therefore is very challenging to extend to dynamic scattering media such as live tissue or fog. Even apparently static media such as white paint are known to evolve over relatively short timescales~\cite{albertazzi2018speckle}.\\
\indent  {The TMs of typical disordered media can also be extremely large, so data-driven approaches seem a reasonable way to tackle the resulting complicated mapping between the input and output patterns.} Machine learning and artificial neural networks (ANNs) have therefore been increasingly applied over the past few years to the problem of classifying objects or imaging through scattering media~\cite{ando2015speckle,satat2017object,turpin2018light,li2018imaging,barbastathis2019use}. These methods however, tend to suffer from the same drawback as the TM approaches, i.e. minimal changes of the scattering medium deteriorate or render the reconstruction impossible. 
Recently, a convolutional neural network {based on the U-net architecture~\cite{Unet}} was shown to be capable of reconstructing the image of a hidden object through optical disorder, while being trained on a similar, but different disorder realizations~\cite{li2018deep}. These results were demonstrated in a very thin diffuser that will therefore exhibit a marked (i.e. wide-angle) ME. A recent extension to the case of multimode fibres was explained as the result of weak but non-zero correlations \cite{bromberg}.\\
\indent In this work we demonstrate the presence of a statistical dependence in the optical field distribution transmitted through a random medium. We purposely consider a system in which no memory effect or any other linear correlations could be present, which is achieved by stacking two glass diffusers at a distance from each other. Numerically modelling allows us to calculate the amount of MI between discretized input and output fields, which gives us an estimate of the input image information carried through this scattering system. We then verify the ability of a U-Net ANN to also image handwritten digits and show that the image reconstruction quality depends on the amount of mutual information. Remarkably, we also find a similar quality image reconstruction in the presence or absence of the memory effect that  underlines the unexpected dominant role played by these statistical dependencies in supervised imaging approaches.
These findings extend the current paradigms for diffuse imaging to random media that are both dynamic and strongly scattering and provide a key to imaging in various scenarios ranging from dynamic multimode fibre endoscopes to imaging through fog and tissue.  \\
{\bf{Random scattering beyond the memory effect.}}
{The scattering configuration we consider is typical to a number of imaging through obscuration experiments,  see Fig.~\ref{fig:one}(a).}
 A collimated laser beam illuminates a spatial light modulator (SLM). The SLM is imaged by a 4f lens system onto a camera. The random scattering medium is made of one or two ground glass diffusers (220 grit, Thorlabs) separated by 5 mm and positioned between the SLM and the first lens of the imaging system.
 \indent One of the main goals of this work is to investigate the underlying physics of imaging through unknown or dynamic scattering media in the absence of the ME, i.e. in regime in which standard approaches based on the measurement of a single transmission matrix or on the autocorrelation of transmitted speckle patterns would fail. The simple {system of two diffusers } excludes the possibility of the ME to influence image reconstruction, which we verified by projecting a simple 3 dot pattern shown in Fig.~\ref{fig:one}(b) and calculating the  autocorrelation, $C(\Delta r) = \langle \int I(r) I(r+\Delta r) /\int I^2(r)\rangle$ of the scattered light far-field  speckle patterns, $I$, (obtained by removing the imaging system) averaged over 600 different diffuser realizations. As we can see in the middle panel of Fig.~\ref{fig:one}(b), the autocorrelation of the light scattered by a single diffuser shows a clear hexagonal pattern (the autocorrelation of the projected pattern), indicating the presence of linear correlations and ME. The right panel shows the same autocorrelation for the two spatially separated diffusers characterised by the absence of any structure, indicating the absence of any ME.\\
{\bf{Statistical dependencies in scattered light.}}
{ In order to track the statistical dependencies between the input and output light intensities for such a system one can build a probability density function (PDF) $P(I(\mathbf{r}_{\text{in}}),I(\mathbf{r}_{\text{out}}))$, where $I(\mathbf{r}_{\text{in}})$ and $I(\mathbf{r}_{\text{out}})$ are the input and output light intensities respectively. This is a very complicated object, since it  captures all the possible combinations of the input and output (continuous) intensities at arbitrary positions in front and behind the scatterer. Assuming the intensity is discretized at $N_i$ levels and there are $N_{\text{in}}$ and $N_{\text{out}}$ observation points in the input and output respectively, the total dimension of this PDF would be $N_i^{N_{\text{in}}}\times N_i^{N_{\text{out}}} $. We numerically study a low-dimensional version of this distribution with $N_i = 2$ and $N_{\text{in}} = N_{\text{out}} = 9$ to get an estimate of the lower bound of the mutual information (MI) between the input and output light distributions. \\  
We modeled the scattering layers by applying a random phase mask to the input field and calculating the output pattern at a particular distance from the incident plane using the standard Fresnel propagation formula (see Supplemental Material \cite{SM} for more model details). The spatial profiles that we use as the phase masks have two characteristic parameters: average width of the features (autocorrelation width) and their average height (standard deviation of the profile). These two parameters determine the strength of the scattering. The input pattern was a set of 9 Gaussian spots, each of their amplitudes being an independent Bernoulli process. The input field therefore  carries exactly 9 bits of spatial information. We calculated the output patterns for each of the 512 combinations of the input spots, each for 1000 different realizations of the random phase masks. We then thresholded the resulting output patterns with respect to the average intensity, selected 9 sub-regions of those binarized patterns and assigned numeric labels to each of the 512 possible combinations of zero or unit intensity within them. The histograms of the pattern labels over the disorder realizations can be interpreted as the conditional distributions of a joint PDF, given a particular input pattern. As all the input patterns are equi-probable, the joint PDF was obtained by simply stacking the conditional PDFs. The MI was then calculated using the standard formula:
\begin{equation}
    \mathcal{I} = H\left[ P(I^{\text{inp}}) \right] - H\left[ P(I^{\text{inp}}|I^{\text{out}})\right].
\end{equation}
The first term is the entropy of the input intensity probability distribution (equal to 9 bits) and the second term is the conditional entropy $H\left[ P(I^{\text{inp}}|I^{\text{out}})\right] = \sum_i p_i H\left[ P(I^{\text{inp}}|I^{\text{out}}=I^{\text{out}}_i) \right]$, where $p_i$ are the probabilities of different output patterns, $H\left[ P(I^{\text{inp}}|I^{\text{out}}=I^{\text{out}}_i) \right]$ are the conditional distributions given a particular output pattern outcome and $H$ is the standard entropy $H\left[ P(x)\right] = -\sum_i P(x_i)\log_2(P(x_i)) $. In order to remove the bias of the conditional entropy estimator, we use its jack-knife version~\cite{paninski2003estimation} (see Supplemental Material \cite{SM} for more details). \\
Figure~\ref{fig2}(a) shows example input and simulated output patterns for 2 different bi-layer phase mask realizations: in the left panel the roughness height standard deviation of the phase masks was 1.3 $\mu$m, while in the right one it was 13 $\mu$m. The broader intensity distribution in the second case reflects the stronger scattering conditions. The resulting simulated PDFs, shown in Fig.~\ref{fig2}(b) indicate that in the stronger scattering scenario (right panel) the resulting  PDF is closer to a simple product of the conditional distributions: some output patterns are more probable than the others, but the conditional distributions given a particular input do not differ significantly. In the weaker scattering case (left panel) the conditional distributions show much more variation, thus one would expect more MI in this case. Indeed the amount of MI vs the separation, $d_o$, of pixels in the 9 spot output pattern is shown in Fig.~\ref{fig2}(c) and indicates that there is always more information in the weak scattering case (left panel) compared to the stronger scattering case (right panel). These graphs also show that the information is not uniformly spread across the speckle image for weaker scattering: more information is contained at { a distance of around 100 $\mu m$ from the center, where the 
output speckle varies more versus dynamic disorder. The uniform value of the MI in the strong scattering case indicates that the speckles now vary uniformly across the camera image plane when the disorder is changed.  \\
}
\begin{figure}[t!]
    \includegraphics[width=\columnwidth]{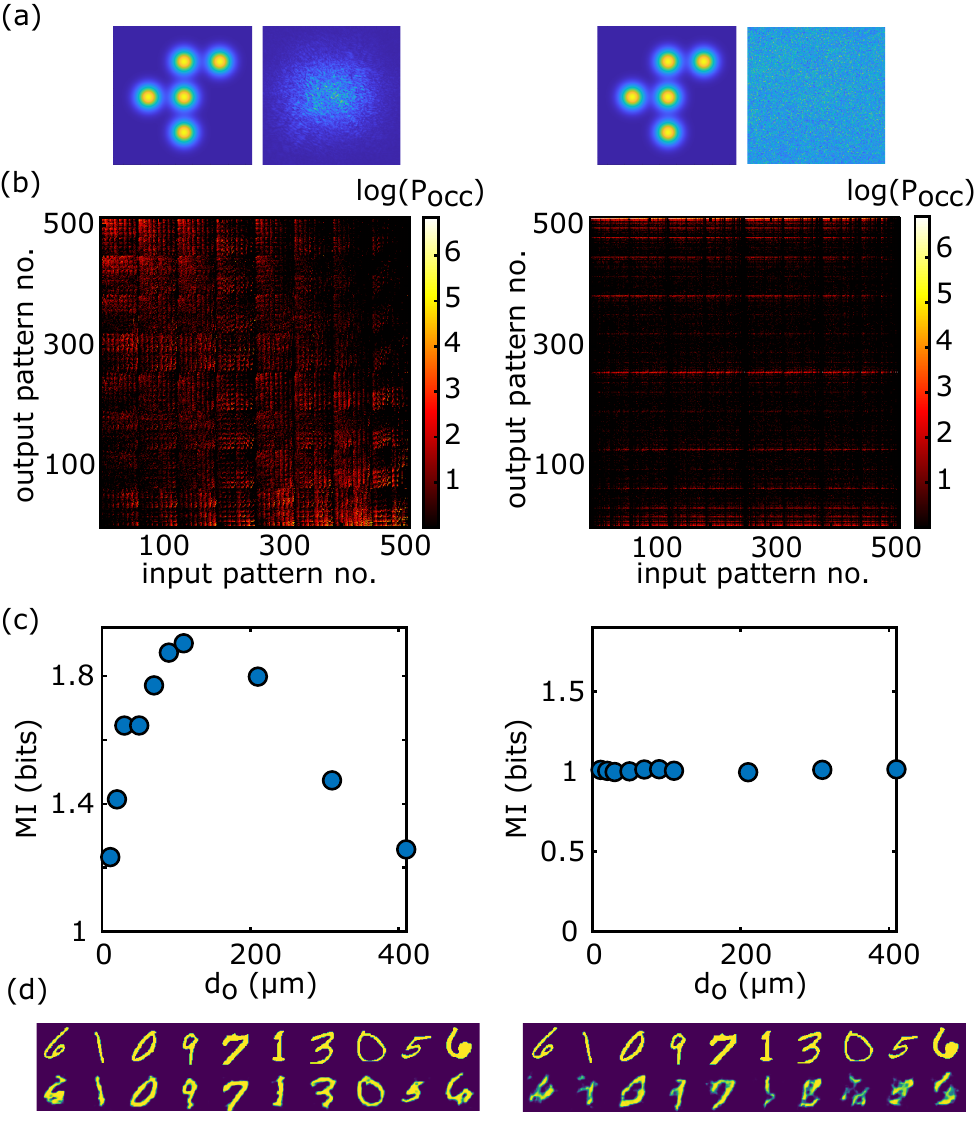}
    \caption{{{\bf {Numerical simulations and mutual information between input and output intensity patterns.}} Left column corresponds to the phase masks with a height standard deviation of 1.3 $\mu$m, right column to 13 $\mu$m. (a) Examples of the input and output simulated patterns used for MI estimation. FWHM of the input spots is 20 $\mu$m. The separation between their centers is 40 $\mu$m. The speckle images have 1$\times$1 mm field of view. (b) Joint PDFs showing the number of occurrences of a particular input and output pattern pair. (c)  MI in bits calculated from the PDFs. $d_o$ is the separation of the pixels in the output pattern see Supplemental Material \cite{SM} for the exact model description. (d) Examples of the image reconstruction through an unseen phase-mask realization: top row shows the input images, the lower row shows the reconstructed images. The reconstruction MSE is 0.059 and 0.078 for the left and right panels respectively. }}
    \label{fig2}
\end{figure}
{\bf{Imaging through scattering media beyond the memory effect.}}
{ The presence of MI however, does not of course guarantee the existence of an easy procedure for its recovery. We show that these statistical dependencies can be picked up by an ANN {that is} trained to transform output speckle patterns into (unseen) input images of objects placed before the scatterer.  
We first verified this on our numerical model. We calculated the speckle patterns in the bi-layer phase mask simulation with the first 1000 MNIST digit images being the input and changing the disorder for each consecutive input image. We repeated this process 32 times using an independent set of phase-masks each time to form a dataset of 32000 speckle/digit-image pairs, which was then used to train a U-net ANN. The imaging performance was tested on a separate set of speckle-image pairs, where neither the input images nor the disorder realization has been used in a training dataset. Examples of the testing results are shown in Fig.~\ref{fig2}(d). The left panel for the weaker scattering case clearly shows the ability to recover unseen images even in the absence of any linear correlations. {{As one might expect from Fig.~\ref{fig2}(c), the stronger scattering case that has a lower MI, also shows worse image reconstruction although, even with just 1 bit of MI, the main features are still recognisable.}}   \\
\begin{figure}[t!]
    \includegraphics[width=8cm]{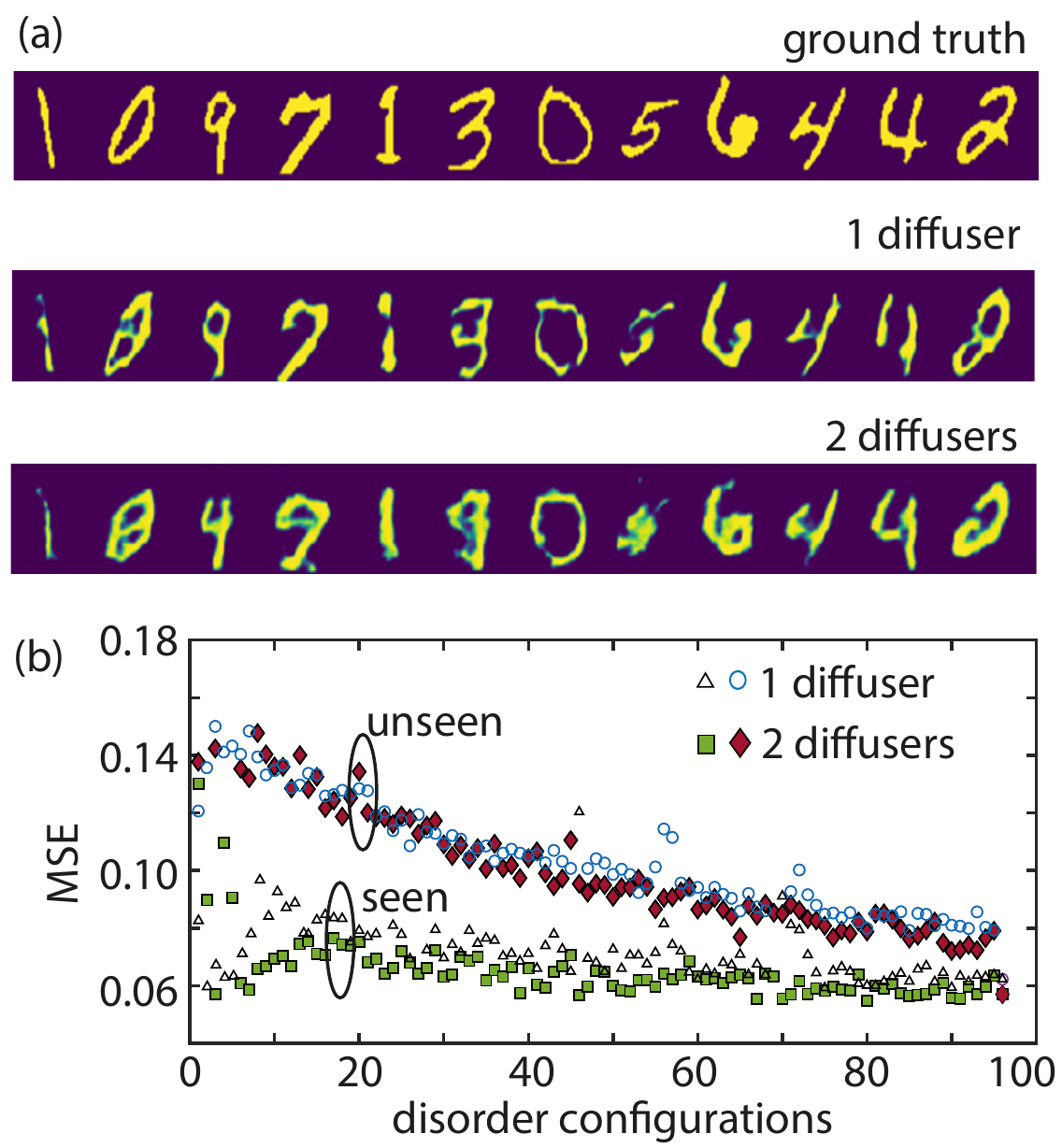}
    \caption{{\bf{Image reconstruction results.}} (a) Examples of ground truth images (unseen during ANN training) together with  image reconstructions (after training on 95 different disorder configurations) with 1 diffuser and two diffusers. (b)  Reconstruction mean-squared error (MSE) for increasing number of disorder configurations used for training, showing how MSE improves continuously with increasing number of training configurations for unseen images.}
    \label{fig:two}
\end{figure} 
\indent We also experimentally verified these findings. We measured the output speckle patterns corresponding to 1000 MNIST digit input images \cite{MNIST}, {with a digital mirror device (DMD) used as an SLM} and repeated this by translating the scattering medium in the transverse plane for 96 different non-overlapping regions. Therefore, each of the 96 repetitions involves completely different microscopic realisations of the random medium, albeit with the same average property, i.e. grit. This data is used to train a U-net ANN \cite{Unet}, following the same architecture explained in detail in Ref.~\cite{li2018deep}.\\
\indent Examples of ANN image reconstruction of digits (unseen during the training) are shown in Fig.~\ref{fig:two}(a). The top row shows ground truth examples with the reconstruction with just one diffuser (middle row) and then for two diffusers (lower row) from which it can be seen that despite the absence of ME, the ANN is still able to reconstruct hidden images successfully. As shown in Fig.~\ref{fig:two}(b) the reconstruction mean squared error continues to decrease with the number of disorder configurations used for training. Moreover, the MSE for both 1 and 2 diffusers decreases at a similar rate and with the same absolute values. This indicates that the ME is never actually playing a major role in the ANN reconstruction, regardless of its presence. This is rather surprising as one might expect a strong linear correlation property to be the dominant feature captured by the high-dimensional interpolation properties of ANNs. Rather, our findings indicate that the ANN is extracting information from the nonlinear dependency and possibly not only here, but also in previous studies that relied on simple single-scattering systems \cite{Unet,tian2}.  \\
{\bf{Conclusions.}}
 We have shown the presence of statistical dependencies {beyond linear correlations} within optical the scattered intensities { by calculating the MI of an input-output probability distribution}  for a system where no linear correlations are possible.  Recent work has also shown how these ANN approaches can be extended to imaging through not only dynamic random media but also at different depths and defocus conditions, thus indicating that these results are not specific to a given imaging system \cite{tian2,bromberg}. \\
\indent Looking forward, a first obvious extension would be to apply these results to imaging through inherently dynamical and changing scattering media such as living tissue or fog. This leads to further questions such as how these results extend to the case in which one physically modifies the microscopic properties, e.g. average scatterer size rather than moving a diffuser  that has fixed statistical properties. There are also implications for applications of scattering media for secure encoding and transmission of information. These systems typically rely on the fact that any given random medium is practically unclonable and therefore acts as one-pad encryption key. However, our results seem to imply that knowledge of the statistical properties of the scattering medium (e.g. the average distribution of refractive index perturbations) is sufficient to decode scrambled information, with important implications on the security of these encoding approaches~\cite{uppu2019asymmetric,fratalocchi}.
{Finally identification of the shape of the nonlinear correlations could provide insight for their in-depth theoretical study in analogy to the theory of linear correlations~\cite{Berkovitsa,Feng1988}}
\\
{\bf{Data availability}} All the data and codes related to this work are available at~\cite{dataset}\\
{\bf{Acknowledgements.}}    We acknowledge discussions with J. Bertolotti.
 This work was supported by the Royal Academy of Engineering, Chair in Emerging Technologies scheme, by the Engineering and Physical Sciences Research Council of the UK (EPSRC) Grant Nos. EP/T00097X/1 and EP/S026444/1 and by the UK MOD University Defence Research Collaboration (UDRC) in Signal Processing.

\end{document}


\title{Supplemental material -- Statistical Dependencies Beyond Linear Correlations in Light Scattered by Disordered Medial}%
\author{Ilya Starshynov}
\author{Alex Turpin}%
\author{Philip Binner}%
\author{Daniele Faccio}%
\affiliation{School of Physics \& Astronomy, University of Glasgow, Glasgow G12 8QQ, United Kingdom }

\date{\today}

\begin{abstract}
This document contains the detailed description of the numerical calculations of the statistical dependencies of the speckle patterns emitted by disordered layers. We also describe  the procedures for obtaining the numeric probability density functions that were used for mutual information calculation. Finally it contains a detailed description of the setup used in our experiment.
\end{abstract}
\maketitle
\section{Description of the numerical model}
\begin{figure}[h!]
    \includegraphics[width=\textwidth]{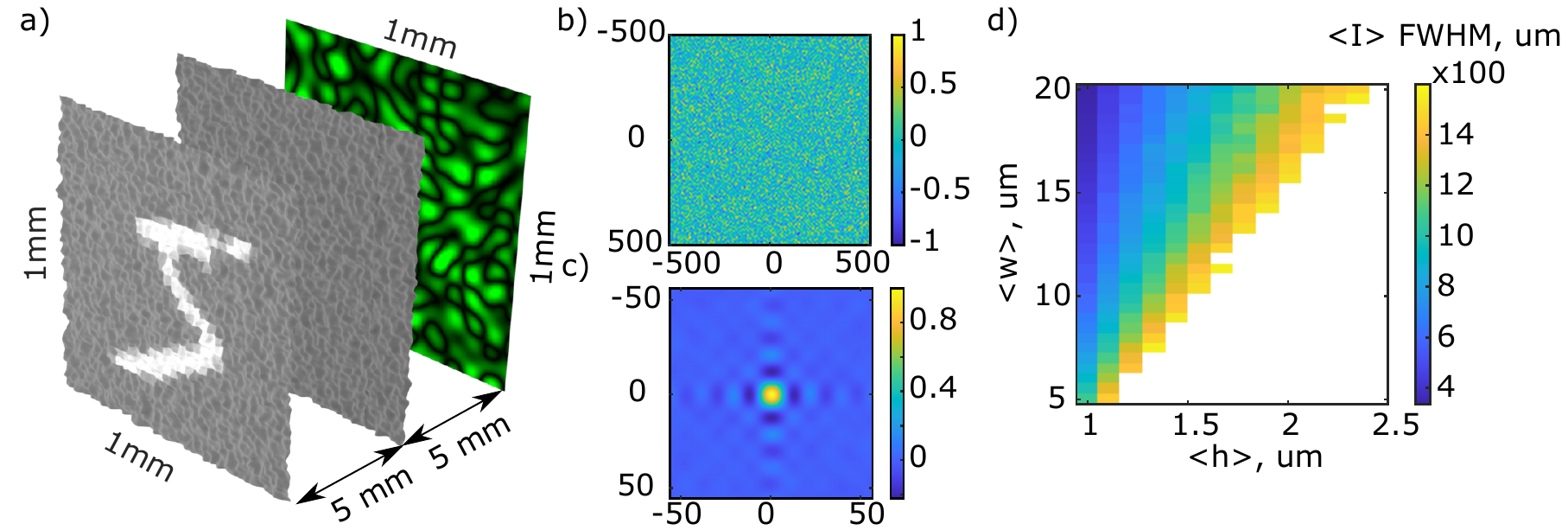}
    \caption { (a) Numerical model setup. (b) Random distribution of the phase-shift across space that serves as a model scattering layer. (c) The atuocorrelation of (b) showing the average width of its features $\langle w \rangle$. The units for both (b) and (c) are microns. (d) Dependence of the average intensity spot size at 5 $mm$ from the scattering layer for a 100 $\mu m$ wide Gaussian input beam on the average width $\langle w \rangle$ and height $\langle h \rangle$ of the phase mask features. }
    \label{fig1}
\end{figure}

The numerical model is described in Fig.~\ref{fig1}a. The incident wavefront with a given intensity distribution, wavelength $\lambda$ = 500 nm and a flat phase is projected onto a random phase mask of dimension 1$\times$1 $mm$. The field at a distance $d$=5 mm is calculated using the Fresnel propagation formula~\cite{born2013principles}. A second and different phase mask is then imposed onto this field and the speckle pattern after another $d$=5 mm of propagation is calculated. The random phase masks are generated by the following procedure: the central $k_{max} \times k_{max}$ square part of a zero-filled array of the size $N_k \times N_k$ is filled with random complex numbers (real and imaginary parts - normally distributed zero mean and unit variance), and an inverse Fourier transform of this array is taken. The result of the inverse Fourier transform resembles a speckle pattern itself and it's real part shown in Fig.~\ref{fig1}b,c is further treated as a spatial distribution of the phase shift thus serving a model of a rough surface. Since the Fourier spectrum of the phase mask spatial features is bounded by $k_{max}$ those features have a fixed average width proportional to $N_{k}/k_{max}$.

The average width of the rough surface features is an important parameter controlling it's scattering angle~\cite{goodman2007speckle, svitasheva2011random}, along with the average surface roughness (standard deviation of the surface height), which can be modelled simply by rescaling the resulting phase mask. The dependence of the scattering angle (average intensity distribution size at a fixed propagation distance) is shown in Fig.~\ref{fig1}d. As expected the average intensity distribution broadens with the increase of the surface roughness or decrease in the surface feature average width, while the average speckle grain size remains almost the same. 

\section{Mutual information between the input and output intensity distributions}
\begin{figure}[t!]
    \includegraphics[width=\textwidth]{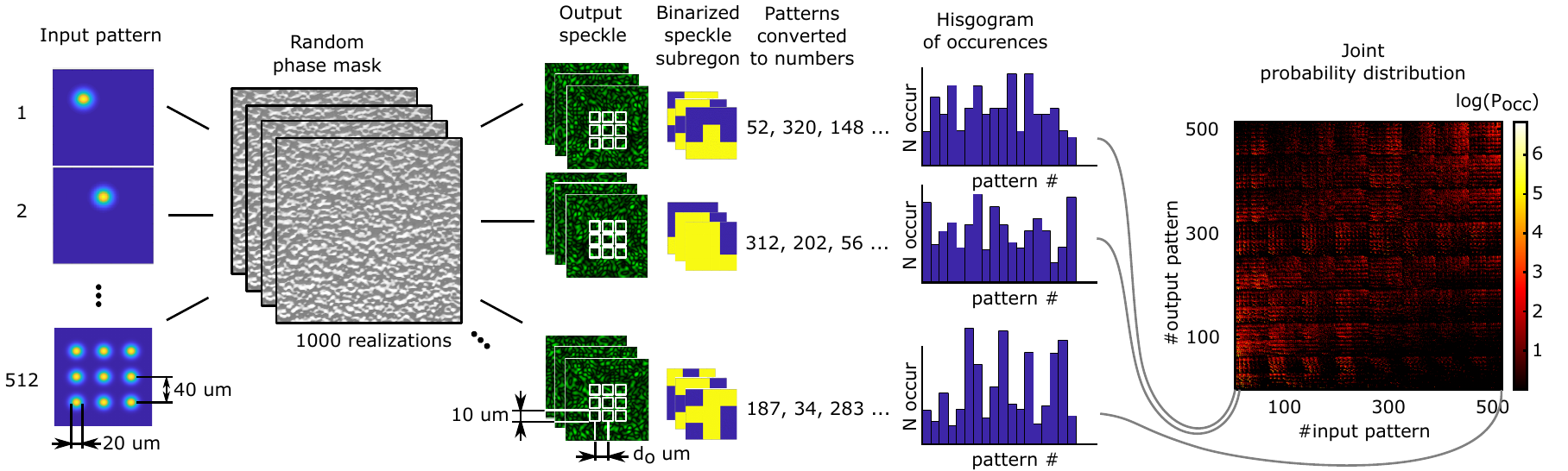}
    \caption{Procedure of construction of the joint probability distribution of the input and output intensity patterns.}
    \label{fig2}
\end{figure} 
Quantification of information transmission from the input to the output of a scattering medium is a challenging task. Since both input and output fields are continuous variables both in space and in the field strength, the dimensionality of the input-output joint probability distribution is not  tractable. Here we estimate a lower bound of the information transfer rate through a set of phase masks by considering a discrete low-dimensional subset of the joint probability space. We assume that the input "image" consists of 9 dots in a square pattern, see Fig.~\ref{fig2}, which can be on or off. There are 2$^9$ = 512 possible configurations of this input pattern. We assume that they are all equiprobable, which means that the input image contains exactly 9 bits of input information. For each configuration of the input pattern we calculate 1000 different output speckle patterns by varying the phase masks between the input and output (but keeping the statistical properties of the phase masks the same). We binarize the output speckles by comparing their local intensity to the disorder average and assigning 0 to the regions below the average and 1 to the regions above it. We then select 3$\times$3 binary patterns, as shown in Fig.~\ref{fig2}b and unwrap them into 9 bit binary numbers. We can vary the separation between the output pattern pixels, $d_o$ thus probing the information content at different regions of the output speckle. Following this procedure, for each input pattern we have a sample of a 1000 output patterns upon varying the disorder. We treat these as conditional distributions of a joint probability distribution, and as all the input variants are equiprobable we can simply stack the conditional distributions into the joint one. Now any statistical property, e.g. mutual information can be calculated from this distribution.

\section{Removing the bias of the MI estimate}
\begin{figure}[b!]
    \includegraphics[width=0.5\textwidth]{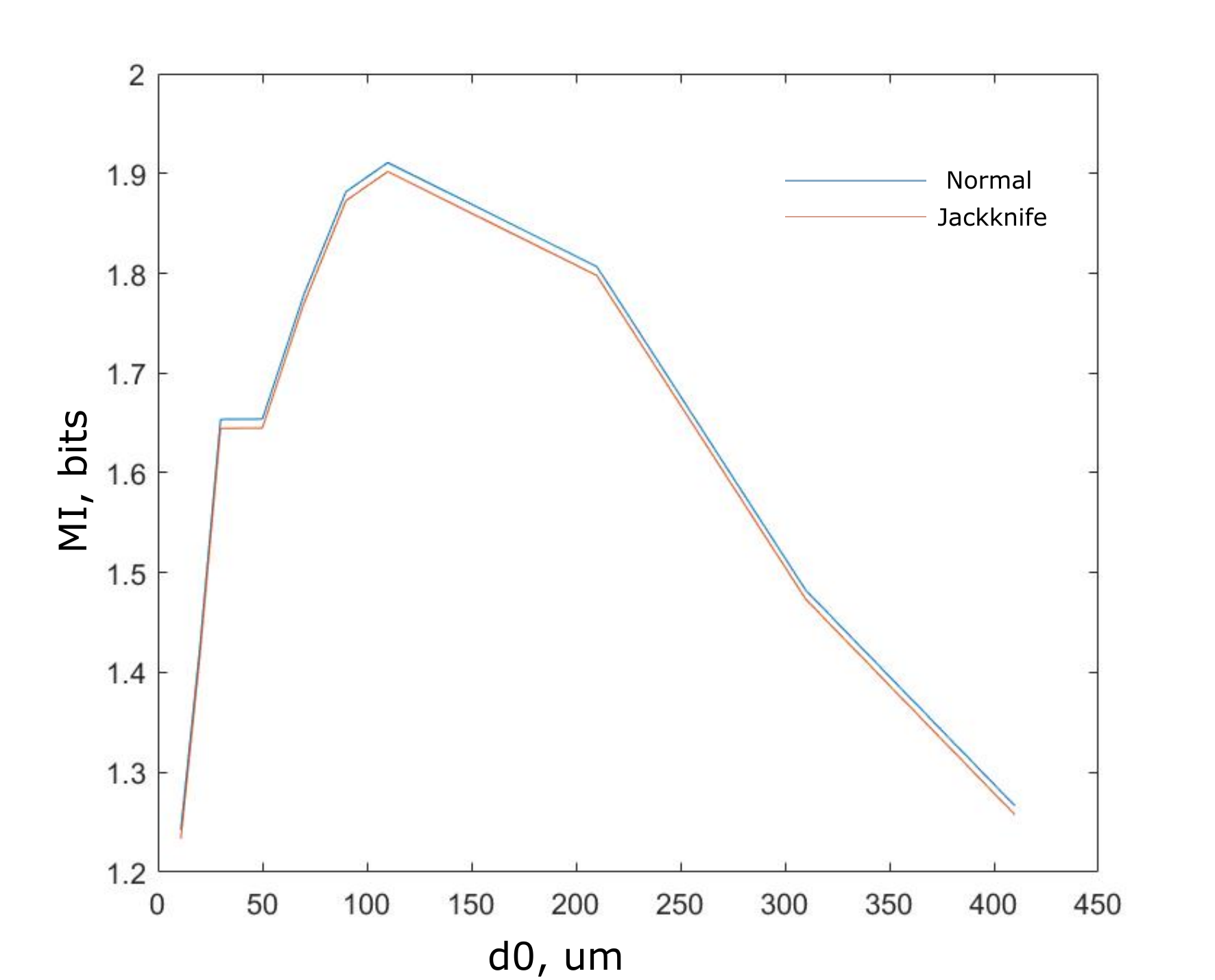}
    \caption{Comparison of the normal (Eq. 1) and the jackknife (Eq. 2) estimators of the conditional entropy for calculating the MI of the input-output PDF for a weaker scattering sample.}
    \label{fig3}
\end{figure}

There are numerous ways of estimating the MI~\cite{paninski2003estimation}. We have chosen  Eq. (1) of the main text because we have prior knowledge of the input probability distribution and thus we would need to estimate only the conditional entropies in the second term. As we have histograms for those, the obvious way is to use the quantity:
\begin{equation}
    H\left[ P(I^{\text{inp}}|I^{\text{out}}=I^{\text{out}}_i) \right] = -\sum_j \mathcal{P}_j \log_2 (\mathcal{P}_j),
\end{equation}
where $\mathcal{P}_j = P_j(I^{\text{inp}}|I^{\text{out}}=I^{\text{out}}_i)$ are simply bin values of the rows of the 2D histogram in Fig 2. It is well known however, that such an estimator is biased, underestimating the entropy and thus leading to an overestimate of the MI. The bias reduces with the number of measurements, but in order to verify that it does not affect our MI estimates we used the jackknife version of this estimator~\cite{efron1981jackknife}:
\begin{equation}
    H_{JK} = N H - \frac{N-1}{N} H^{N-1},
\end{equation}
where $N$ is the number of samples, $H$ is the usual estimator from Eq. (1) and $H^{N-1}$ is the average of the estimator (1) using all N but one samples, over all $N$ possible of the sample that is being skipped. The dependence of the MI on $d_o$ for the 1.3 $\mu m$ scattering layers obtained using normal and jackknife estimators is shown in Fig.~\ref{fig3}. As can be seen the difference between these two values is small compared to the values themselves, thus telling us that the bias does not significantly affect the estimates in our results. 

\section{Experimental setup and data processing}

\begin{figure}[h!]
    \includegraphics[width=0.8\textwidth]{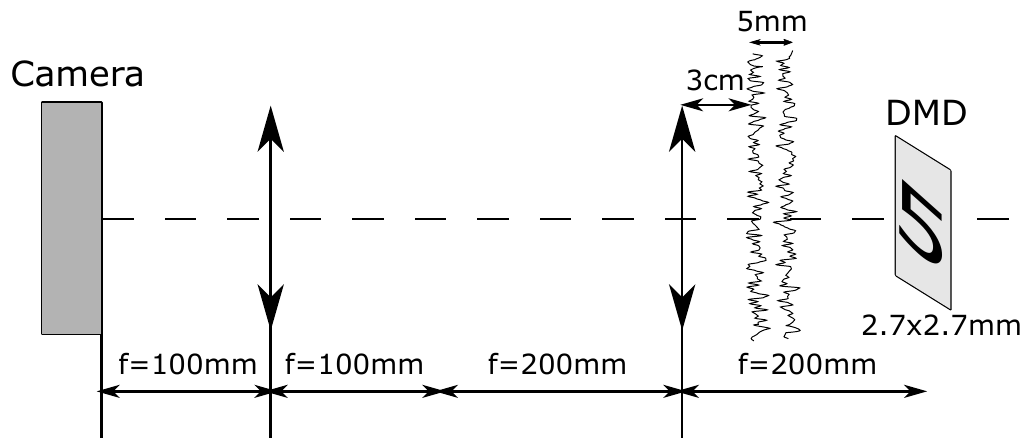}
    \caption{Experimental setup}
    \label{fig4}
\end{figure}

Experimental setup is illustrated inf Fig.~\ref{fig4}. A laser beam (DJ532-40 Laser diode, Thorlabs) which is expanded to 5mm diameter is directed onto a 200x200 pixel area (2.7x2.7mm) on a DMD (DLP7000, Texas instruments) where MNIST digits are displayed. The DMD is imaged onto the camera with a 4f imaging system with a magnification of 2. The scattering object composed of 2 glass diffusers (DG100X100-220, Thorlabs) separated by 5mm is placed at around 3cm from the input lens of the 4f system. The scattering medium can be translated (perpendicularly to the optical axis) by an XY-stage. A sequence consisting of the first 1000 images of the MNIST handwritten digit dataset is displayed on the DMD with the camera capturing the resulting speckle patterns on a hardware trigger. This process is repeated for 96 positions of the diffusers chosen in such a way that the displacement between the adjacent regions of the diffusers is greater than the illuminated spot, thus probing completely different disorder realizations. The speckle images are cropped to 256x256 pixels with one speckle occupying around 3pixels. The first 900 speckle-input image pairs from the first 95 illuminations spots are used to train the U-net ANN, which is then tested on the previously unseen 100 digits passed through an unseen disorder realization.